\documentclass[journal]{IEEEtran}
\ifCLASSINFOpdf
\else
\fi
\usepackage[dvips]{graphicx}
\usepackage{epstopdf}
\usepackage{subfigure}
\usepackage{amsthm,amssymb}
\usepackage{amsmath}
\usepackage{algpseudocode}
\usepackage{algorithm,cite}
\usepackage{enumerate}
\usepackage{fixltx2e}
\usepackage{mathtools}
\usepackage[utf8]{inputenc}
\usepackage[T1]{fontenc}

\begin{document}
\makeatletter
\def\ps@IEEEtitlepagestyle{
  \def\@oddfoot{\mycopyrightnotice}
  \def\@evenfoot{}
}
\def\mycopyrightnotice{
  {\footnotesize
  \begin{minipage}{\textwidth}
  \copyright~2018 IEEE. Personal use of this material is permitted. Permission from IEEE must be obtained for all other uses, in any current or future media, including reprinting/republishing this material for advertising or promotional purposes, creating new collective works, for resale or redistribution to servers or lists, or reuse of any copyrighted component of this work in other works.
  \end{minipage}
  }
}

\title{Monopulse beam synthesis\\using a sparse single-layer of weights}

\author{Semin Kwak, Joohwan Chun and Sung Hyuck Ye
\thanks{S. Kwak is with \'Ecole polytechnique f\'ed\'erale de Lausanne (EPFL), EPFL 1015 Lausanne, Switzerland. J. Chun is with
Korea Advanced Institute of Science and Technology (KAIST), Daejeon 305-701 Korea, and
S. Ye is with Agency for Defense Development (ADD), Yuseong PO Box 35-71, Daejeon 305-600, Korea.
This work was done while S. Kwak was with KAIST and was supported in part by ADD, under contract ADD-140101.}}

\maketitle
\begin{abstract}

A conventional monopulse radar system uses three beams; sum beam, elevation difference beam and azimuth difference beam,
which require different layers of weights to synthesize each beam independently.
Since the multi-layer structure increases hardware complexity, many simplified structures based on a single layer of weights
have been suggested.
In this work, we introduce a new technique for finding disjoint and fully covering sets of weight vectors, each of which
constitutes a sparse subarray, forming a single beam.
Our algorithm decomposes the original non-convex optimization problem for finding disjoint weight vectors
into a sequence of convex problems.
We demonstrate the convergence of the algorithm
and show that the interleaved array structure is able to meet difficult beam constraints.

\end{abstract}

\begin{IEEEkeywords}
Monopulse radar, sparse array, interleaved array, convex optimization, argumentative reselection algorithm, alternating projection method.
\end{IEEEkeywords}

\IEEEpeerreviewmaketitle

\section{Introduction}
A monopulse radar with an antenna array needs multiple beams; the sum beam and the delta beams,
on a same antenna-array face. This, in turn, requires multiple layers of weights {\it i.e.,}
transmit-receive modules (TRMs) to shape each beam, independently and optimally.
However, it is costly and structurally complicated to attach multiple TRMs on each antenna.
Therefore, many researchers have engaged in the problem
of subarraying and assigning a single weight on each antenna
heuristically~\cite{Nickel1995} and systematically
~\cite{rocca2009hybrid, haupt2005interleaved, lopez2001subarray, d2007effective, manica2009fast, omt2008, Caoris, MDUrso}.

\begin{figure}[t]
\centering
\subfigure[An example of clustered non-overlapping subarrays, which form a single-layer of weights.
This particular structure is sometimes called the {\it common weight} array.
In general, the summation of the antenna response is carried out
before the analog-digital conversion (ADC) and the summation of the subarray responses, after the ADC.]
{\includegraphics[width=0.95\columnwidth]{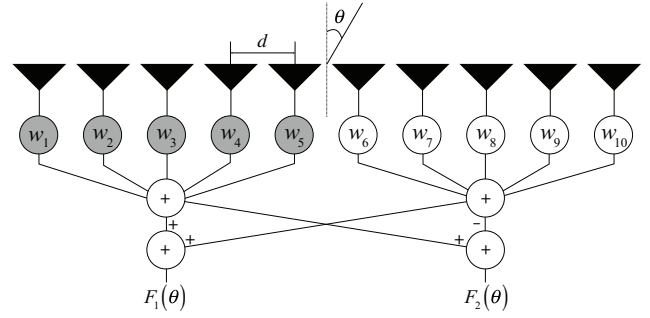}}
\subfigure[An example of fully interleaved structure which form a single-layer of weights with sparse and disjoint subarrays.
The summation of the antenna responses is carried out before the ADC.]
{\includegraphics[width=0.95\columnwidth]{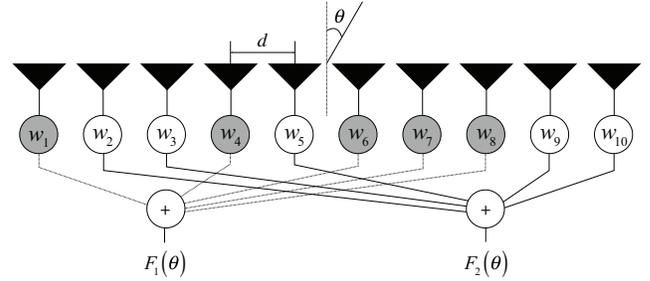}}
\caption{Two types of single-layer monopulse antenna structure.}\label{fig:seperateSubarray}
\end{figure}

The first approach
is to use
multiple and clustered sub-arrays,
whose responses are combined to obtain multiple beams~\cite{lopez2001subarray, manica2009fast,d2007effective}.
For example, Figure~\ref{fig:seperateSubarray}(a) shows two clustered non-overlapping subarrays,
which form a single-layer of weights, that are used to obtain the sum beam $F_1$ and the delta beam $F_2$.
Each subarray response is generated by combining the antenna responses in analog manner,
and each beam response is obtained by combining the subarray responses in digital manner.
Figure~\ref{fig:seperateSubarray}(a) shows {\it non-overlapping} subarrays~\cite{kwak2016},
but {\it partially overlapping}~\cite{Morabito2010, mohammed2017synthesizing} structures have been studied as well.

The second approach
is to use sparse and irregular subarrays.
As an example,
the fully interleaved thinned linear array (FITLA) structure~\cite{haupt2005interleaved}
is shown in Figure~\ref{fig:seperateSubarray}(b).
The larger aperture of each sparse sub-array,
compared with that of a dense array with the same number of weights,
gives sharper and narrower beam while the irregularity of the sparse array suppresses grating lobes.

Above mentioned beamforming structures can be synthesized
by formulating
a constrained minimization problem, which is in general, non-convex.
Although this minimization can be carried out with an inefficient global optimization method, two
ingeniously crafted methods, the
alternating projection method and
the hybrid method exist.

The alternating projection method~\cite{omt2008, bfmp, tmga, ombucci, han, Leonardo} iteratively finds
an intersecting point of two sets $\mathcal{M}$ and $\mathcal{B}$,
where $\mathcal{M}$ specifies the excitation constraints and $\mathcal{B}$, the beam pattern,
by repeatedly projecting the point in a current set onto the other.
A difficulty with this method is that one of the two sets, in general, is not convex,
and therefore the starting point must be chosen carefully to ensure the convergence
to the global minimum.
We remark that the alternating projection method appears in
a variety of algorithms, sometimes disguisedly, including the direction of arrival finding algorithm~\cite{ziskind} and
the alternating direction implicit (ADI) method for solving partial differential equations~\cite{varga}.

The hybrid method~\cite{lopez2001subarray, manica2009fast, rocca2009hybrid, d2007effective, omt2008, Caoris, MDUrso, Isernia} also
decomposes the constrained minimization problem
into two; one a convex problem and the other, usually a non-convex problem.
This method has been successfully applied to the monopulse beam synthesis by iteratively finding
the weights as well as subarray grouping for the structures in Figures 1(a) and 1(b)
~\cite{rocca2009hybrid, haupt2005interleaved, lopez2001subarray, d2007effective, manica2009fast, omt2008, Caoris, MDUrso}.

In general, however, both the alternating projection method and the hybrid method
involve a non-convex optimization step, and therefore, require a global optimization algorithm~\cite{bfmp}
or a good choice of the starting point~\cite{MDUrso}.

We propose a new algorithm which solves the original non-convex problem
for finding the structure in Figure 1(b)
by decomposing it into
a sequence of
l1-minimization problems, which are convex.
In a sense, the proposed algorithm is a variant of the alternating projection method, where the
non-convex constraint set is replaced with a more convenient convex set~\cite{quijano}.

\section{Data model}\label{Section:MAS}
Assuming omni-directional antennas, let us
define the array response vector by
${\bf a}(\theta) = [1, e^{-jkd sin\theta}, \cdots , e^{-jkd(N-1)sin \theta}]^T$,
where $j= \sqrt{-1}$, and $k = 2 \pi / \lambda$. The angle $\theta$ denotes the bearing of a target
and the constant $d$, antenna spacing.
Then sum beam response $F_1$ and delta beam response $F_2$ are respectively, expressed as
\begin{equation}{F_1}\left( {\theta} \right) = {\bf{a}}{\left( \theta \right)^H}{{\bf{M}}^H}{{\bf{w}}_1},
\quad {\bf w}_1 \in {\mathbb C}^{N \times 1} \label{eqn:sum_vect_form},
\end{equation}
\begin{equation}
{F_2 }\left( \theta  \right) = {\bf{a}}{\left( \theta \right)^H}{{\bf{M}}^H}{{\bf{w}}_2},
\quad {\bf w}_2 \in {\mathbb C}^{N \times 1}, \label{eqn:diff_vect_form}
\end{equation}
where ${\bf{M}}$ represents the mutual coupling matrix.
The vectors ${\bf w}_1$ and ${\bf w}_2$ are {\it disjoint} and fully {\it covering} weight vectors.
For example, for the array in Figure~\ref{fig:seperateSubarray}(b), we have
\begin{eqnarray}
{\bf w}_1 &=& [w_1, 0, 0, w_4, 0, w_6, w_7, w_8, 0, 0]^T, \\
{\bf w}_2 &=& [0, w_2, w_3, 0, w_5, 0, 0, 0, w_9, w_{10}]^T.
\end{eqnarray}
%
The role of weight vectors ${\bf w}_1$
and ${\bf w}_2$ is to compensate the mutual coupling as well as to shape the beams
under the altered array response vector, ${\bf M}{\bf a}(\theta)$.

Let $\Theta_1$ and $\Theta_2$ be the sets of side-lobe angles of the sum beam and the delta beam, respectively,
and $\theta_0$, the bore-sight angle.
A monopulse radar functions properly, if $F_1$ and $F_2$ are synthesized to satisfy each beam constraints, {\it i.e.},
the weight vectors ${\bf w}_1$ and ${\bf w}_2$
belong to the sets defined by
\begin{equation}
{\mathcal{C}_1 } = \left\{ {{{\bf{w}}_1 }\left| {{F_1 }\left( {{\theta _0}} \right)
= \mu,\;{{\left| {{F_1 }\left( {{\theta_{1,m_1}}} \right)} \right|}^2} \le {\tau _1}} \right.} \right\},
\label{eqn:feasibleSet:sum}
\end{equation}
\begin{equation}
{\mathcal{C}_2 } = \left\{ {{{\bf{w}}_2 }\left|
{\begin{array}{*{20}{c}}
{{F_2 }\left( {{\theta _0}} \right) = 0,\;
{{\left| {{F_2 }\left( {{\theta_{2,m_2}}} \right)} \right|}^2} \le {\tau _2},}\\
{{\left. \frac{\partial F_2 \left( \theta  \right)} {{\partial \theta }} \right|}_{\theta  = \theta _0}
= {{\mu s}}}
\end{array}} \right.} \right\},
\label{eqn:feasibleSet:del}
\end{equation}
where
$\{ \theta_{1,m_1}\}_{m_1=1}^{M_1} \in \Theta_1$ and $\{ \theta_{2,m_2} \}_{m_2=1}^{M_2} \in \Theta_2$
indicate sampling points in the side-lobe regions.
The constants $M_1$ and $M_2$ represent the number of samples in the side-lobe regions for each beam.
Only one sample at $\theta_0$ is taken in the main-lobe regions.
Here, $\mu$ is defined as the array gain for the sum beam at the bore sight.
The bounds $\tau_1$ and $\tau_2$ denote the maximum side-lobe levels (SLLs) of $F_1$ and $F_2$, respectively,
and the constant $s$ is the slope of $F_2$ at the bore sight.
The sets ${\mathcal C}_1$ and ${\mathcal C}_2$ are convex sets, since $F_1$ and
$F_2$ are linear functions of ${\bf w}_1$ and ${\bf w}_2$, respectively.
See Equation (\ref{eqn:sum_vect_form}) and (\ref{eqn:diff_vect_form}).

\section{Argumentative reselection algorithm}
To find disjoint weight vectors such as
${{\bf{w}}_1}$ and ${{\bf{w}}_2}$ in Figure~\ref{fig:seperateSubarray}(b),
we shall build an optimization problem and propose an algorithm to solve the problem.
We call the process, argumentative reselection algorithm,
since the process is comparable to the situation
where many people argue for their individual benefit, but eventually
reach a compromise with which all can accept.
\subsection{The problem}
Now let us define the two-variable cost function 
${\tilde J} \left( {{{\bf{w}}_1},{{\bf{w}}_2}} \right) \buildrel \Delta \over =
{{\left| {{{\bf{w}}_1}} \right|}^T} \cdot {\left| {{{\bf{w}}_2}} \right|}$,
where 
${\left| \cdot \right|}$
takes the
element-wise absolute value.
Then the problem is to find ${\bf\hat w}_1$ and ${\bf\hat w}_2$ such that
\begin{equation}\label{opt:OP}
\begin{aligned}
\left( {\bf\hat w}_1,{\bf\hat w}_2 \right) =& \mathop {{\mathop{\rm argmin}\nolimits}}\limits_{{\bf w}_1, {\bf w}_2}
{\tilde J}\left( {\bf w}_1,{\bf w}_2 \right)\\
& {\rm subject}\;{\rm to}\;{{\bf{w}}_1} \in {\mathcal{C}_1},{{\bf{w}}_2} \in {\mathcal{C}_2}.
\end{aligned}
\end{equation}
The disjoint requirement of ${\bf w}_1$ and ${\bf w}_2$ is built-into the cost function
because if we are able to minimize the cost of ${\tilde J}$ down to zero, then we shall obtain a disjoint pair ${\bf w}_1 \in {\mathcal C}_1$
and ${\bf w}_2 \in {\mathcal C}_2$.
Otherwise, there are no disjoint ${\bf w}_1 \in {\mathcal C}_1$ and ${\bf w}_2 \in {\mathcal C}_2$,
and we need to relax the specifications in ${\mathcal C}_1$ and ${\mathcal C}_2$.
This trial and error approach of the parameter selection is quite common for beam synthesis problems~\cite{omt2008}.

\subsection{The algorithm}\label{Section:DRA}
\begin{figure}[t]
\centering
\subfigure[Magnitude distributions of $\left|{\bf{w}}_1\right|$ and $\left|{\bf{w}}_2\right|$ after the first iteration.]
{\includegraphics[width=0.95\columnwidth]{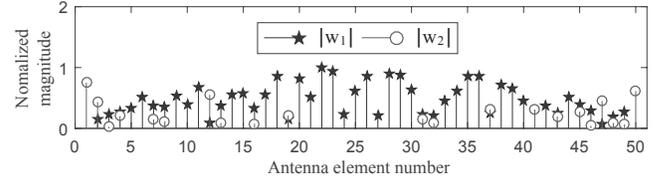}}
\subfigure[Magnitude distributions of $\left|{\bf{w}}_1\right|$ and $\left|{\bf{w}}_2\right|$ after the second iteration.]
{\includegraphics[width=0.95\columnwidth]{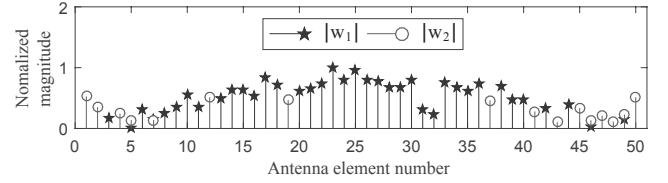}}
\subfigure[Magnitude distributions of $\left|{\bf{w}}_1\right|$ and $\left|{\bf{w}}_2\right|$ after the fourth iteration.]
{\includegraphics[width=0.95\columnwidth]{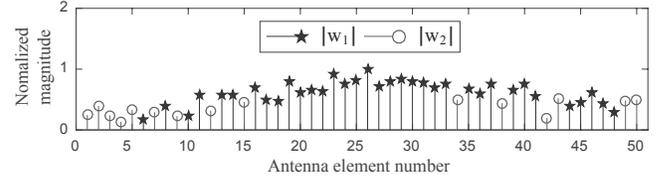}}
\caption{Antenna weight distributions as the argumentative reselection algorithm proceeds.}
\label{fig:DRalgorithm}
\end{figure}

A difficulty in the above optimization problem is that the two-variable cost function is not convex~\cite{Boyd2004}.
However, if one of the weight vectors ${\bf w}_1$ and ${\bf w}_2$ is constant, then the resulting
one-variable cost function $J(\cdot)$ becomes convex.
This leads us to build a new algorithm for Problem (\ref{opt:OP}), which utilizes {\it penalizing} vectors
${\bf p}_1$ and ${\bf p}_2$: \\
\\
Choose random ${\bf w}_1$ and ${\bf w}_2$;
\newline {\bf repeat} 
\begin{enumerate}[(i)]
\item Set ${\bf p}_1 \coloneqq \left|{{\bf{w}}}_2\right|$
and solve \\
${{{\bf{w}}}_1 } \coloneqq \mathop {{\mathop{\rm argmin}\nolimits} }\limits_{{{\bf{w}}_1 }}
J({\bf w}_1, {\bf p}_1)
\;{\rm s.t.}\;{{\bf{w}}_1 } \in {\mathcal{C}_1 }$;. \label{stepC}
\item Set ${\bf p}_2 \coloneqq\left|{{\bf{w}}}_1\right|$
and solve \\
${{{\bf{w}}}_2 } \coloneqq \mathop {{\mathop{\rm argmin}\nolimits} }\limits_{{{\bf{w}}_2 }}
J({\bf w}_2, {\bf p}_2)
\;{\rm s.t.}\;{{\bf{w}}_2 } \in {\mathcal{C}_2 }$; \label{stepD}
\end{enumerate}
{\bf until} converge
%
%
\subsubsection{Intuitive analysis of the algorithm}
Let us consider the statement (ii) in the above algorithm,
where the penalizing vector ${\bf p}_2 \buildrel \Delta \over = [p_1, p_2, \cdots , p_N]^T$
is the result ${\bf w}_1$ after (i) in the current iteration,
and ${\bf w}_2 \buildrel \Delta \over = [w_1, w_2, \cdots , w_N]^T$ is the result after (ii) in the previous iteration.
From
\begin{eqnarray}
J({\bf w}_2, {\bf p}_2) = |w_1| \cdot p_1 + |w_2| \cdot p_2 + \cdots + |w_N| \cdot p_N,
\end{eqnarray}
the new weight ${\bf w}_2$ is to be found by minimizing $J({\bf w}_2, {\bf p}_2)$,
or in other words, by taking a smaller $w_i$
for a larger $p_i$.
Therefore, the vectors ${\bf w}_1$ and ${\bf w}_2$ tend to become disjoint as the iteration continues.

Figure~\ref{fig:DRalgorithm} shows an example of antenna distributions
in the course of
the argumentative reselection process.
The initial values for the elements of 
${\bf w}_1$ and ${\bf w}_2$ are all $1$s.
After the first iteration (See Figure~\ref{fig:DRalgorithm}(a).),
there are $18$ shared antennas by the sum beam and the delta beam.
However, as the algorithm proceeds, the shared antennas are removed gradually
(See Figure~\ref{fig:DRalgorithm}(b).), and then completely (See Figure~\ref{fig:DRalgorithm}(c).).

\subsubsection{Argumentative reselection algorithm for multiple beams}
In the case of $K$ beams, {\it i.e.}, $K$ weight vectors in the sets $\{ {\mathcal C}_k \}_{k=1}^K$,
the $K$-variable cost function of the optimization problem is defined as:
\begin{equation}\label{generalJ}
{\tilde J}\left( {{{\bf{w}}_1},{{\bf{w}}_2}, \cdots ,{{\bf{w}}_K}} \right) = {1 \over 2}\sum\limits_{\scriptstyle i,j = 1, \hfill \atop
  \scriptstyle i \ne j \hfill} ^K {{{\left| {{{\bf{w}}_i}} \right|}^T} \cdot \left| {{{\bf{w}}_j}} \right|}.
\end{equation}
If we define the penalizing vectors and the one-variable cost function, respectively as:
\begin{equation}\label{generalP}
    {{\bf{p}}_k} =  \sum\limits_{\scriptstyle j = 1, \hfill \atop
  \scriptstyle j \ne k \hfill} ^K {\left| {{{{{\bf w}}}_j}} \right|}, \quad
  J({\bf w}_k, {\bf p}_k) = |{\bf w}_k|^T \cdot {\bf p}_k,
\end{equation}
then 
we have Algorithm 1 below.
\begin{algorithm}
\caption{Algorithm for the argumentative reselection process}\label{prop_algo1}
{\bf initialize} ${{{\bf{w}}}_k} \coloneqq {{\bf{randn}}(N,1)+j{\bf{randn}}(N,1)}\;
       {\rm for}\;k \in \left\{ {1, \ldots ,K} \right\}$, and set ${J^{(0)}} \coloneqq$ inf \\
{\bf for} {$l = 1,2, \cdots$} \\ 
\hspace*{0.3in} {\bf for} {$k = 1:K$} \\
\hspace*{0.6in} ${{\bf{p}}_k} \coloneqq  \sum\limits_{\scriptstyle j = 1, \hfill \atop
              \scriptstyle j \ne k \hfill} ^K {\left| {{{{{\bf w}}}_j}} \right|}$; \\
\hspace*{0.6in} ${{{\bf{ {w}}}}_k} \coloneqq \mathop {{\mathop{\rm argmin}\nolimits} }\limits_{{{\bf{w}}_k}}
              J({\bf w}_k, {\bf p}_k),\;{\rm subject}\;{\rm to}\;{{\bf{w}}_k} \in {\mathcal{C}_k}$; \\
\hspace*{0.3in} {\bf end} \\
\hspace*{0.3in}${J^{(l)}}\coloneqq {\tilde J}\left( {{{{\bf{w}}}_1},{{{\bf{w}}}_2}, \cdots ,{{{\bf{w}}}_K} } \right)$; \\
\hspace*{0.3in} {\bf if} $J^{(l-1)} - J^{(l)} < \epsilon$ {\bf exit} \\
{\bf end} \\
{\bf return} ${{\bf{ \hat{w}}}_k} \coloneqq {{\bf{ {w}}}_k}\;{\rm for}\;k \in \left\{ {1, \ldots ,K} \right\};$
\end{algorithm}


The above stated algorithm with two weight vectors
is a special case of Algorithm 1, when $K=2$.
Now the convergence of Algorithm 1 is proven below. \\

\noindent {\bf Theorem}: {The sequence $J^{(l)}$ in Algorithm 1 is monotonically
decreasing and bounded below by zero, and thus convergent.}
\begin{proof}
Writing the intermediate results explicitly, let
${\bf w}_k^{(l)}$ be the optimal vector obtained after the $k$th inner-iteration of the $l$th outer-iteration,
and we define
\begin{equation}
J_k^{(l)} \buildrel \Delta \over = {\tilde J}({\bf w}_1^{(l)}, {\bf w}_2^{(l)}, \cdots , {\bf w}_{k}^{(l)}, {\bf w}_{k+1}^{(l-1)}, \cdots, {\bf w}_K^{(l-1)}).
\end{equation}
Then when $k=1$,
\begin{eqnarray}
{{\bf w}}_1^{\left( {l + 1} \right)} &=& \mathop {{\mathop{\rm argmin}\nolimits} }\limits_{{{\bf{w}}_1}} J({\bf w}_1, {\bf p}_1) \\
&=& \mathop {{\mathop{\rm argmin}\nolimits} }\limits_{{{\bf{w}}_1}}
{\left| {{{\bf{w}}_1}} \right|^T} \cdot \left(  \sum\limits_{\scriptstyle j = 2} ^K
{\left| {{{\bf w}}_j^{\left( l \right)}} \right|} \right).
\end{eqnarray}
Therefore,
\begin{align}
J_1^{\left( {l + 1} \right)} &= {\left| {{{\bf w}}_1^{\left( {l + 1} \right)}} \right|^T}
\!\cdot\! \left( {\sum\limits_{\scriptstyle j = 2} ^K {\left| {{{\bf w}}_j^{\left( l \right)}} \right|} } \right)
\!+\! {1 \over 2}\sum\limits_{\scriptstyle i,j = 2, \hfill \atop
  \scriptstyle i \ne j \hfill} ^K {{{\left| {{{\bf w}}_i^{\left( l \right)}} \right|}^T}
  \!\cdot\! \left| {{{\bf w}}_j^{\left( l \right)}} \right|}\\
&\le {\left| {{{\bf w}}_1^{\left( l \right)}} \right|^T}
\cdot \left( {\sum\limits_{\scriptstyle j = 2} ^K {\left| {{{\bf w}}_j^{\left( l \right)}} \right|} } \right) + {1 \over 2}\sum\limits_{\scriptstyle i,j = 2, \hfill \atop
  \scriptstyle i \ne j \hfill} ^K {{{\left| {{{\bf w}}_i^{\left( l \right)}} \right|}^T}
  \cdot \left| {{{\bf w}}_j^{\left( l \right)}} \right|}  \\
 &= J_K^{\left( l \right)},
\end{align}
\normalsize
so that $J_K^{(l)} \ge J_1^{(l+1)}$.
Similarly $J_k^{(l+1)} \ge J_{k+1}^{(l+1)}$, and therefore
$J_K^{(l)} \ge J_K^{(l+1)} \ge J_K^{(l+2)} \ge \cdots$.
Now the proof is complete because $J_K^{(l)} = J^{(l)}, \forall l$, and $J^{(l)}$ are bounded below by zero
from the definition of ${\tilde{\it J}}$ in Equation (\ref{generalJ}).
\end{proof}

\section{Numerical results}\label{Section:NR}

We use
the MOSEK solver which uses an interior point method~\cite{Boyd2004} 
to solve the convex optimization problem.
The simulation is performed with MATLAB under the hardware condition of i7-4790-3.6GHz (CPU) and 16GB RAM.

To demonstrate the applicability of the proposed algorithm under mutual coupling, we shall consider an idealized coupling
matrix~\cite{friedlander1991direction}:
%
\begin{equation}
{\bf{M}} = \left[ {\begin{array}{*{20}{c}}
1&\rho &{{\rho ^2}}& \cdots &{{\rho ^{N - 1}}}\\
\rho &1&\rho & \ddots & \vdots \\
 {{\rho ^2}}&\rho &1& \ddots &{{\rho ^2}}\\
 \vdots & \ddots & \ddots & \ddots &\rho \\
{{\rho ^{N - 1}}}& \cdots &{{\rho ^2}}&\rho &1
\end{array}} \right]
\end{equation}
Namely, we approximate ${\bf M}$ with the covariance matrix of an autoregressive process of order 1.
According to our simulation study, the proposed algorithm appears to be robust under different coefficient values of $\rho$.
However, a proper coupling matrix must be determined experimentally for each particular antenna array before
the algorithm is applied.

First, let us consider a uniform linear array of $N = 120$ antennas with $d=\lambda/2$, and therefore,
of an aperture size, $59.5 \lambda$.
We assume the mutual coupling constant $\rho$ to be 0.1.
The sidelobe regions for the sum and delta beams are respectively
$\Theta_1 = [1^{\circ}, 90^{\circ}] \cup [-90^{\circ}, -1^{\circ}]$ and $\Theta_2 = [1.2^{\circ}, 90^{\circ}] \cup [-90^{\circ}, -1.2^{\circ}]$.
The maximum SLLs are assumed to be $\tau_1 = \tau_2 = -16.7 {\rm dB}$, and the slope,
$s=-100 deg^{-1}$.
Letting $\mu=1$, we calculate the exact value of maximum SLLs.
However,
the value of the parameter $\mu$ is not important, since the maximum SLLs are relative value (in ${\rm dB}$) of $\mu$.
We have taken 1001 samples distributed evenly in $[-90^{\circ} , 90^{\circ}]$.

\begin{figure}[t]
\centering
\subfigure[Normalized magnitudes of weights using the proposed algorithm]
{\includegraphics[width=0.95\columnwidth]{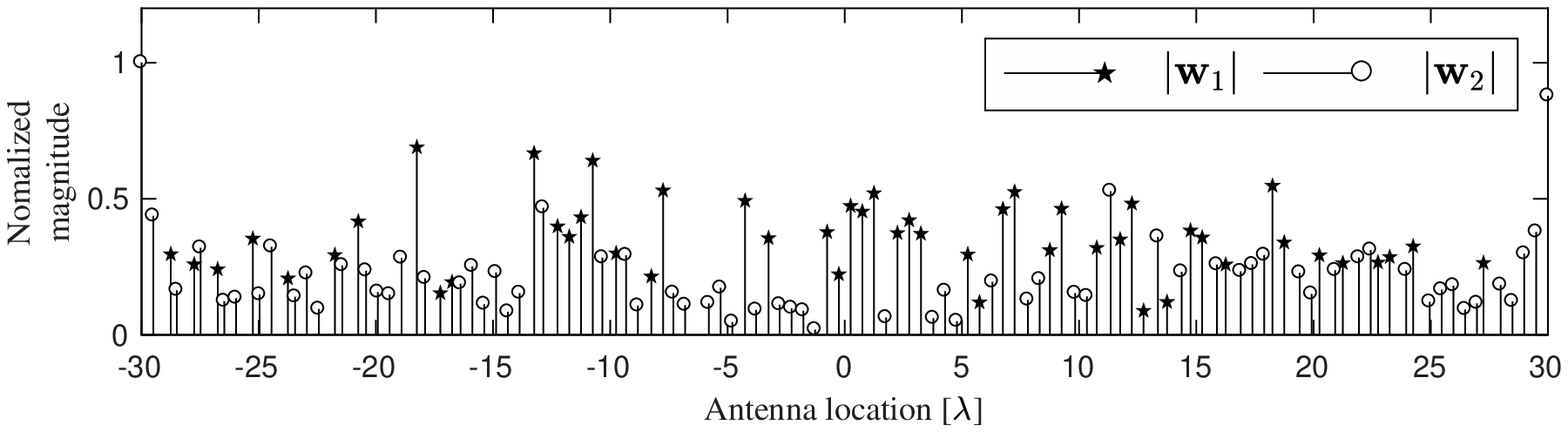}}
\subfigure[Normalized magnitudes of weights using the common shared weight approach (${\bf{w}}_c$: common weight vector).]
{\includegraphics[width=0.95\columnwidth]{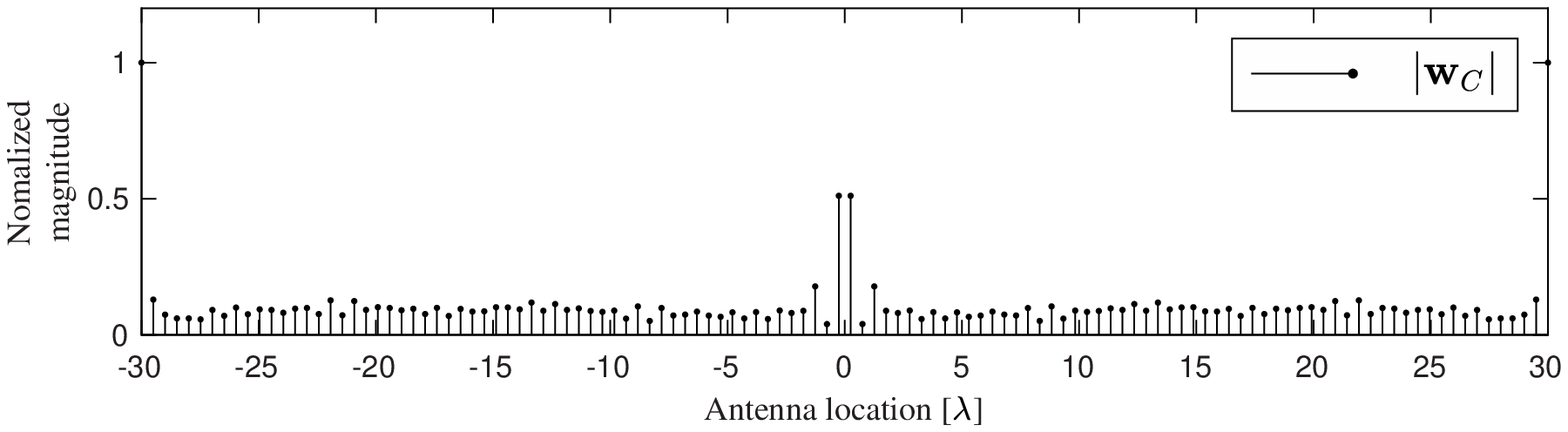}}
\subfigure[Beam patterns with the weight vectors in (a).]
{\includegraphics[width=0.45\columnwidth]{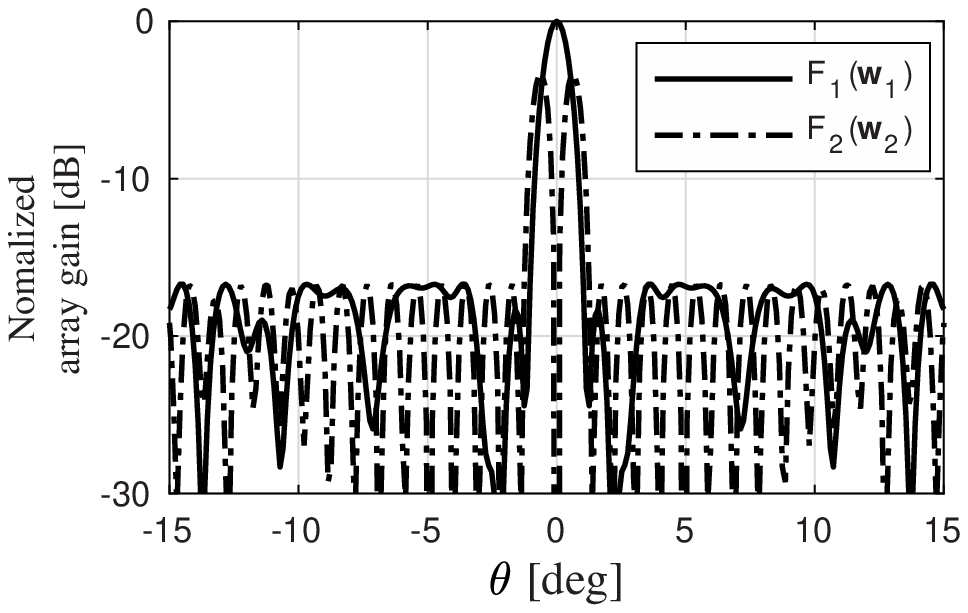}}
\subfigure[Beam patterns with the weight vectors in (b).]
{\includegraphics[width=0.45\columnwidth]{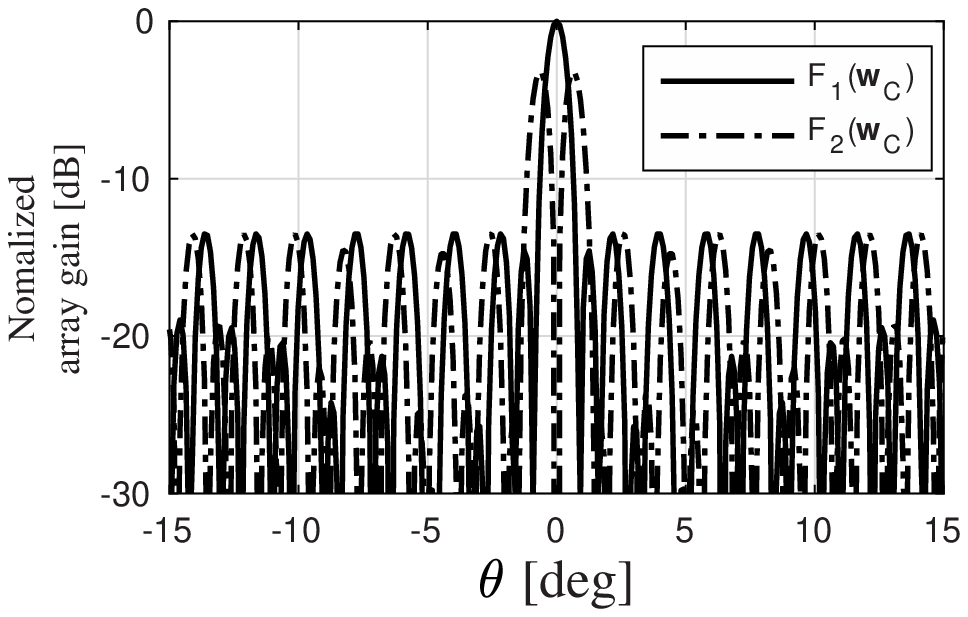}}
\subfigure[Success rate of finding a feasible solution for different maximum sidelobe levels.]
{\includegraphics[width=0.95\columnwidth]{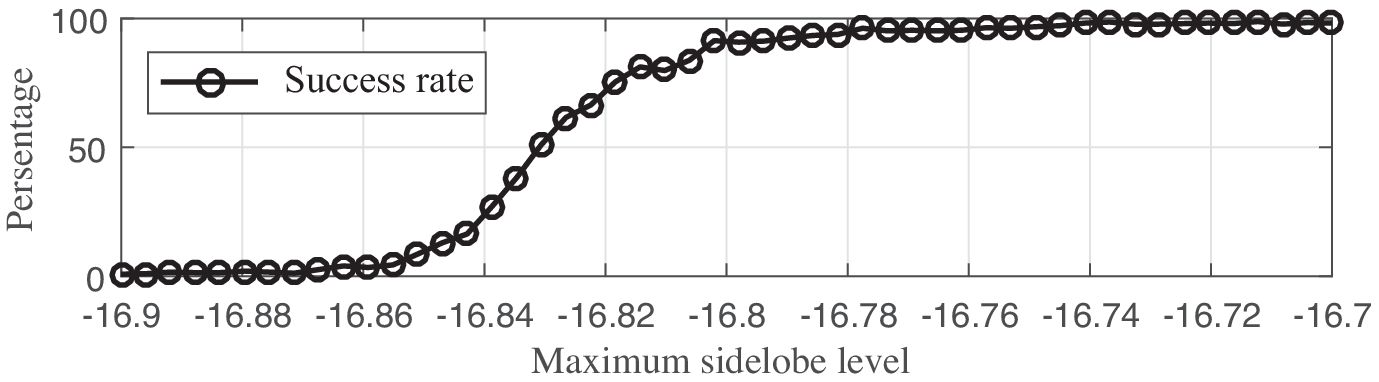}}
\caption{Comparison of the proposed weights and the common weights.}
\label{img:compare}
\end{figure}

Algorithm 1 with the constant $\epsilon = 10^{-5}$
successfully finds a pair of disjoint weight vectors;
53 weights for $F_1$ and 67 weights for $F_2$ (Figure~\ref{img:compare}(a)),
using the MOSEK solver. The computation time is $6.12$ second.
The corresponding beam patterns
(Figure~\ref{img:compare}(c)) meet the specifications we set above. Here, the 3${\rm dB}$ beam width is $0.99^{\circ}$.

On the other hand,
the non-overlapping common weight approaches~\cite{kwak2016} 
could not find a feasible solution satisfying the specifications,
(even when the mutual coupling matrix is the identity).
If we relax the SLL requirements
until the remaining beam requirements as well as the initial settings are met,
then a solution could be found as shown in Figure~\ref{img:compare}(b).
The corresponding beam patterns have higher SLLs as shown in Figure~\ref{img:compare}(d). The maximum SLLs of both
the sum beam and difference beam are $-13.01{\rm dB}$.
However, it must be stressed that our simulation study is limited, and we cannot conclude that the proposed algorithm is
superior to the general common weight algorithms~\cite{Morabito2010, kwak2016, mohammed2017synthesizing}.

To verify the reliability of the argumentative reselection algorithm,
we examine the success rate of finding a feasible solution
through Monte Carlo simulation.
We count successful runs out of $500$ trials with
random initial penalizing vectors $\bf{p}_1$,
for each SLL from $-16.9{\rm dB}$ to $-16.7{\rm dB}$.
As shown in Figure~\ref{img:compare}(e),
the proposed algorithm converges to zero
with a $96.2\%$ rate
when the SLL is above $-16.78{\rm dB}$.

\begin{figure}[t]
\centering
\subfigure[Disjoint antenna distributions of three beams $F_1$, $F_2$ and $F_3$.]
{\includegraphics[width=0.8\columnwidth]{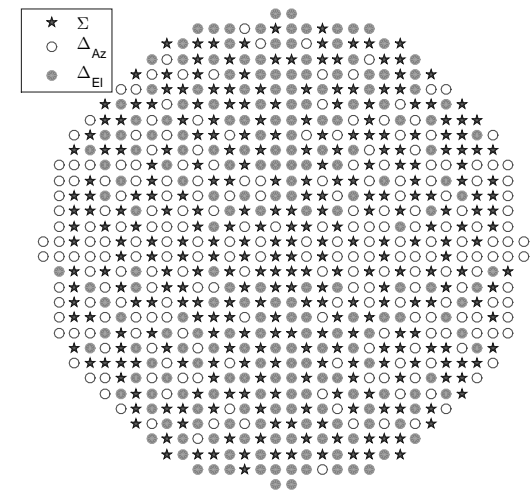}}
\subfigure[Sum beam pattern $F_1$.]
{\includegraphics[width=0.8\columnwidth]{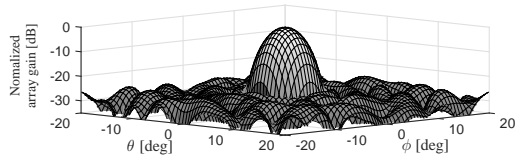}}
\subfigure[Difference beam pattern for azimuth $F_2$.]
{\includegraphics[width=0.8\columnwidth]{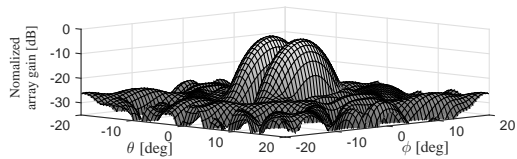}}
\subfigure[Difference beam pattern for elevation $F_3$.]
{\includegraphics[width=0.8\columnwidth]{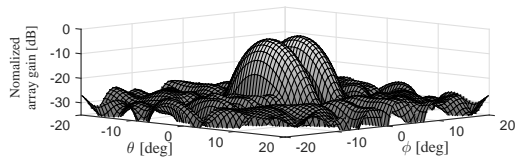}}
\subfigure[Convergence of the cost function $J$ and the number of shared antennas during the iteration.]
{\includegraphics[width=0.8\columnwidth]{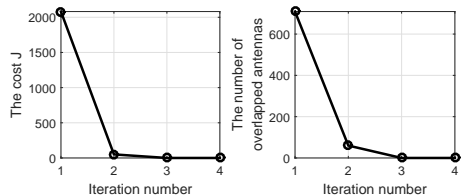}}
\caption{Two-dimensional monopulse radar beams with disjoint weights.}
\label{img:2D}
\end{figure}

Next, we consider a 2D planar array of 756 antennas with $d = \lambda/2$, which forms
three beams; the sum beam $F_1$, the azimuth difference beam $F_2$ and the elevation difference beam $F_3$.
We ignore the mutual coupling effect for simple exposition.
Our beam specifications for $F_2$ and $F_3$ will be the same, and therefore, we shall
describe $F_1$ and $F_2$ only.
%
The side-lobe regions for $F_1$ and $F_2$ are
$\Theta_1 = [5^{\circ}, 90^{\circ}] \cup [-90^{\circ}, -5^{\circ}]$ and
$\Theta_2 = [8^{\circ}, 90^{\circ}] \cup [-90^{\circ}, -8^{\circ}]$, respectively.
The maximum SLLs are assumed to be $\tau_1 = \tau_2 = -25 {\rm dB}$, and the slope $s$,
$-22 deg^{-1}$.

To speed up the computation of Algorithm 1, we divide the side-lobe regions into two,
region {\it a} and region {\it b}:
\begin{eqnarray}
\Theta_{1a} &=& [5^{\circ}, 20^{\circ}) \cup (-20^{\circ}, -5^{\circ}],\\
\Theta_{1b} &=& [20^{\circ}, 90^{\circ}] \cup [-90^{\circ}, -20^{\circ}],\\
\Theta_{2a} &=& [8^{\circ}, 20^{\circ}) \cup (-20^{\circ}, -8^{\circ}],\\
\Theta_{2b} &=& [20^{\circ}, 90^{\circ}] \cup [-90^{\circ}, -20^{\circ}],
\end{eqnarray}
Then we take $10 \times 2$ evenly spaced samples each, from $\Theta_{1a}$ and $\Theta_{2a}$,
and $35 \times 2$ evenly spaced samples each, from
$\Theta_{1b}$ and $\Theta_{2b}$.
Again Algorithm 1 with $\epsilon = 10^{-5}$
finds three disjoint weight vectors; 301 weights for $F_1$, 233 weights for $F_2$ and 222 weights for $F_3$.
See Figure~\ref{img:2D}(a). The computation time is $33.89$ minutes.
The corresponding beam patterns, Figure~\ref{img:2D}(b), (c), and (d)
meet the given specifications,
and the convergence is shown in
Figure~\ref{img:2D}(e). The $3{\rm dB}$ beamwidth of the sum beam is $4.2^{\circ}$ for both azimuth and elevation angles.

\section{Concluding Remarks}\label{Section:Conclusions}
We have presented the argumentative reselection algorithm which partitions an antenna array into sparse sets,
such that the sets of weights are disjoint and give independent desired beam patterns.
Sparse subarrays with disjoint weights have a more degree of freedom, and therefore
have better control on the beamshapes compared to common ({\it i.e.,} shared) weight structure.
As future work,
it may be considered to include the crossing counts of the beamforming network into the objective function
to reduce the complexity of the feeder structure.

\bibliographystyle{IEEEtran}
\bibliography{references}

\begin{thebibliography}{10}
\providecommand{\url}[1]{#1}
\csname url@samestyle\endcsname
\providecommand{\newblock}{\relax}
\providecommand{\bibinfo}[2]{#2}
\providecommand{\BIBentrySTDinterwordspacing}{\spaceskip=0pt\relax}
\providecommand{\BIBentryALTinterwordstretchfactor}{4}
\providecommand{\BIBentryALTinterwordspacing}{\spaceskip=\fontdimen2\font plus
\BIBentryALTinterwordstretchfactor\fontdimen3\font minus
  \fontdimen4\font\relax}
\providecommand{\BIBforeignlanguage}[2]{{%
\expandafter\ifx\csname l@#1\endcsname\relax
\typeout{** WARNING: IEEEtran.bst: No hyphenation pattern has been}%
\typeout{** loaded for the language `#1'. Using the pattern for}%
\typeout{** the default language instead.}%
\else
\language=\csname l@#1\endcsname
\fi
#2}}
\providecommand{\BIBdecl}{\relax}
\BIBdecl

\bibitem{Nickel1995}
U.~Nickel, ``Subarray configurations for digital beamforming with low sidelobes
  and adaptive interference suppression,'' in \emph{Proceedings of IEEE
  International Radar Conference}, Alexandria, VA, USA, May 1995, pp. 714--719.

\bibitem{rocca2009hybrid}
P.~Rocca, L.~Manica, R.~Azaro, and A.~Massa, ``A hybrid approach to the
  synthesis of subarrayed monopulse linear arrays,'' \emph{IEEE Transactions on
  Antennas and Propagation}, vol.~57, no.~1, pp. 280--283, 2009.

\bibitem{haupt2005interleaved}
R.~Haupt, ``Interleaved thinned linear arrays,'' \emph{IEEE transactions on
  antennas and propagation}, vol.~53, no.~9, pp. 2858--2864, 2005.

\bibitem{lopez2001subarray}
P.~Lopez, J.~Rodriguez, F.~Ares, and E.~Moreno, ``Subarray weighting for the
  difference patterns of monopulse antennas: joint optimization of subarray
  configurations and weights,'' \emph{IEEE Transactions on Antennas and
  Propagation}, vol.~49, no.~11, pp. 1606--1608, 2001.

\bibitem{d2007effective}
M.~D'Urso, T.~Isernia, and E.~F. Meliado, ``An effective hybrid approach for
  the optimal synthesis of monopulse antennas,'' \emph{IEEE transactions on
  antennas and propagation}, vol.~55, no.~4, pp. 1059--1066, 2007.

\bibitem{manica2009fast}
L.~Manica, P.~Rocca, M.~Benedetti, and A.~Massa, ``A fast graph-searching
  algorithm enabling the efficient synthesis of sub-arrayed planar monopulse
  antennas,'' \emph{IEEE Transactions on Antennas and Propagation}, vol.~57,
  no.~3, pp. 652--663, 2009.

\bibitem{omt2008}
O.~M. Bucci, M.~D'Urso, and T.~Isernia, ``Some facts and challegenes in array
  antenna synthesis problems,'' \emph{Automatika}, vol.~49, no. 1-2, pp.
  13--20, 2008.

\bibitem{Caoris}
S.~Caorsi, A.~Massa, M.~Pastorino, and A.~Randazzo, ``Optimization of the
  difference patterns for monopulse antennas by a hybrid real/integer-coded
  differential evolution method,'' \emph{IEEE Transactions on Antenna and
  Propagation}, vol.~53, no.~1, pp. 372--376, Jan 2005.

\bibitem{MDUrso}
M.~D'Urso and T.~Isernia, ``Solving some array synthesis problems by means of
  an effective hybrid approach,'' \emph{IEEE Transactions on Antenna and
  Propagation}, vol.~55, no.~3, pp. 750--759, Mar 2007.

\bibitem{kwak2016}
S.~Kwak, J.~Chun, D.~Park, Y.~Ko, and B.~Cho, ``Asymmetric sum and difference
  beam pattern synthesis with a common weight vector,'' \emph{IEEE Antennas and
  Wireless Propagation Letters}, vol.~15, pp. 1622--1625, 2016.

\bibitem{Morabito2010}
A.~F. Morabito and P.~Rocca, ``Optimal synthesis of sum and difference patterns
  with arbitrary sidelobes subject to common excitations constraints,''
  \emph{IEEE Antennas and Wireless Propagation Letters}, vol.~9, pp. 623--626,
  2010.

\bibitem{mohammed2017synthesizing}
J.~R. Mohammed, ``Synthesizing sum and difference patterns with low complexity
  feeding network by sharing element excitations,'' \emph{International Journal
  of Antennas and Propagation}, vol. 2017, 2017.

\bibitem{bfmp}
O.~M. Bucci, G.~Franceschetti, G.~Mazzarella, and G.~Panariello, ``Intersection
  approach to array pattern synthesis,'' \emph{IEE Proceedings}, vol. 137,
  no.~6, pp. 349--356, Dec 1990.

\bibitem{tmga}
D.~Trincia, L.~Marcaccioli, R.~Gatti, and R.~Sorrentino, ``Modified projection
  method for array pattern synthesis,'' in \emph{34th European Microwave
  Conferencen}, 2004, pp. 1379--1400.

\bibitem{ombucci}
O.~M. Bucci, D.~Giuseppe, G.~Mazzarella, and G.~Panariello, ``Antenna pattern
  synthesis: A new general approach,'' \emph{Proceesings of the IEEE}, vol.~82,
  no.~3, pp. 358--371, Mar 1994.

\bibitem{han}
Y.~Han, C.~Wan, W.~Sheng, B.~Tian, and H.~Yang, ``Array synthesis using
  weighted alernating projection and proximal splitting,'' \emph{IEEE Antenna
  and Wireless Propagation Letters}, vol.~14, pp. 1006--1009, 2015.

\bibitem{Leonardo}
J.~Leonardo, A.~Quijano, and G.~Vecchi, ``Alternating adaptive projections in
  antenna synthesis,'' \emph{IEEE Transactions on Antenna and Propagation},
  vol.~58, no.~3, pp. 727--737, Mar 2010.

\bibitem{ziskind}
I.~Ziskind and M.~Wax, ``Maximum likelihood localization of multiple sources by
  alternating projection,'' \emph{IEEE Transactions on acoustics, speech and
  signal processing}, vol.~36, no.~10, pp. 1553--1560, Oct 1988.

\bibitem{varga}
R.~Varga, \emph{Matrix Iterative Analysis}.\hskip 1em plus 0.5em minus
  0.4em\relax Englewood Cliffs: Prentice-Hall, 1962.

\bibitem{Isernia}
T.~Isernia, F.~Ares~Pena, O.~M. Bucci, M.~D'Urso, J.~Gomez, and J.~Rodriguez,
  ``A hybrid approach for the optimal synthesis of pencil beams through array
  antennas,'' \emph{IEEE Transactions on Antenna and Propagation}, vol.~52,
  no.~11, pp. 2912--2918, Nov 2004.

\bibitem{quijano}
J.~Quijano and G.~Vecchi, ``Alternating adaptive projections in antenna
  synthesis,'' \emph{IEEE Transactions on Antenna and Propagation}, vol.~58,
  no.~3, pp. 727--737, Mar 2010.

\bibitem{Boyd2004}
S.~Boyd and L.~Vandenberghe, \emph{Convex Optimization}.\hskip 1em plus 0.5em
  minus 0.4em\relax Cambridge: Cambridge University Press, 2004, vol.~25.

\bibitem{friedlander1991direction}
B.~Friedlander and A.~Weiss, ``Direction finding in the presence of mutual
  coupling,'' \emph{IEEE transactions on antennas and propagation}, vol.~39,
  no.~3, pp. 273--284, 1991.

\end{thebibliography}

\end{document}